\PassOptionsToPackage{unicode}{hyperref}
\PassOptionsToPackage{hyphens}{url}
\PassOptionsToPackage{dvipsnames,svgnames,x11names}{xcolor}
\documentclass[
]{article}
\usepackage{amsmath,amssymb}
\usepackage{lmodern}
\usepackage{iftex}
\ifPDFTeX
  \usepackage[T1]{fontenc}
  \usepackage[utf8]{inputenc}
  \usepackage{textcomp} 
\else 
  \usepackage{unicode-math}
  \defaultfontfeatures{Scale=MatchLowercase}
  \defaultfontfeatures[\rmfamily]{Ligatures=TeX,Scale=1}
\fi
\IfFileExists{upquote.sty}{\usepackage{upquote}}{}
\IfFileExists{microtype.sty}{
  \usepackage[]{microtype}
  \UseMicrotypeSet[protrusion]{basicmath} 
}{}
\makeatletter
\@ifundefined{KOMAClassName}{
  \IfFileExists{parskip.sty}{%
    \usepackage{parskip}
  }{
    \setlength{\parindent}{0pt}
    \setlength{\parskip}{6pt plus 2pt minus 1pt}}
}{
  \KOMAoptions{parskip=half}}
\makeatother
\usepackage{xcolor}
\usepackage{graphicx}
\makeatletter
\def\maxwidth{\ifdim\Gin@nat@width>\linewidth\linewidth\else\Gin@nat@width\fi}
\def\maxheight{\ifdim\Gin@nat@height>\textheight\textheight\else\Gin@nat@height\fi}
\makeatother
\setkeys{Gin}{width=\maxwidth,height=\maxheight,keepaspectratio}
\makeatletter
\def\fps@figure{htbp}
\makeatother
\providecommand{\tightlist}{%
  \setlength{\itemsep}{0pt}\setlength{\parskip}{0pt}}
\newlength{\cslhangindent}
\newlength{\csllabelwidth}
\newlength{\cslentryspacingunit} 
\newenvironment{CSLReferences}[2] 
 {
  \setlength{\parindent}{0pt}
  \ifodd #1
  \let\oldpar\par
  \def\par{\hangindent=\cslhangindent\oldpar}
  \fi
  \setlength{\parskip}{#2\cslentryspacingunit}
 }%
 {}
\usepackage{calc}

\ifLuaTeX
\usepackage[bidi=basic]{babel}
\else
\usepackage[bidi=default]{babel}
\fi
\babelprovide[main,import]{american}

\def\languageshorthands#1{}
\ifLuaTeX
  \usepackage{selnolig}  
\fi
\IfFileExists{bookmark.sty}{\usepackage{bookmark}}{\usepackage{hyperref}}
\IfFileExists{xurl.sty}{\usepackage{xurl}}{} 
\hypersetup{
  pdftitle={scida: scalable analysis for scientific big data},
  pdfauthor={Chris Byrohl, Dylan Nelson},
  pdflang={en-US},
  colorlinks=true,
  linkcolor={Maroon},
  filecolor={Maroon},
  citecolor={Blue},
  urlcolor={Blue},
  pdfcreator={LaTeX via pandoc}}

\title{scida: scalable analysis for scientific big data}

\usepackage[affil-it]{authblk}
\usepackage{orcidlink}
\author[1%
  ]{Chris Byrohl%
    \,\orcidlink{0000-0002-0885-8090}\,%
    }
\author[1%
  ]{Dylan Nelson%
    \,\orcidlink{0000-0001-8421-5890}\,%
    }

\affil[1]{Heidelberg University, Institute for Theoretical Astronomy,
Albert-Ueberle-Str. 2, 69120 Heideberg, Germany}
\date{15 Sep 2023}

\begin{document}
\maketitle

\hypertarget{summary}{%
\section{Summary}\label{summary}}

\begin{figure}
\centering
\includegraphics{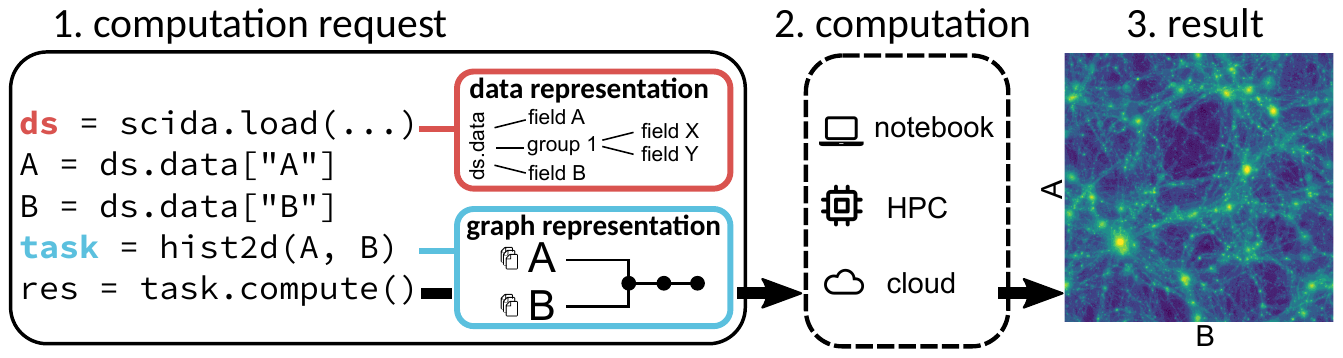}
\caption{Schematic of the workflow. In a user session, a recipe
(i.e.~sequence of analysis operations) for desired data product can be
built by consecutive chaining of operations, which are internally
represented by dask task graphs. Calculation is triggered by the
compute() command, evaluating the graph on a target resource. The
result, much smaller than the original data, is sent back to the user
session for further analysis/plotting. \label{fig:sketch}}
\end{figure}

\textbf{scida} (\url{https://scida.io}) is a Python package for reading and analyzing large
scientific data sets. Data access is provided through a hierarchical
dictionary-like data structure after a simple load() function. Using the
dask library for scalable, parallel and out-of-core computation
(\protect\hyperlink{ref-Dask}{Dask Development Team, 2016}), all
computation requests from a user session are first collected in a task
graph. Arbitrary custom analysis, as well as all available dask (array)
operations, can be performed. The subsequent computation is executed
only upon request, on a target resource (e.g.~a HPC cluster, see Figure
\ref{fig:sketch}).

\hypertarget{features}{%
\section{Features}\label{features}}

Scida begins by providing a clean, abstract, dictionary-like interface
to the underlying data, regardless of its file format or structure on
disk. Physical units are automatically inferred, when possible, and
attached to the dataset. Symbolic and automatic unit conversion is
provided by the pint package (\protect\hyperlink{ref-Pint}{Grecco \&
others, 2012}). This metadata can also be specified by the user via
customizable configuration files.

Scida attempts to automatically determine the type of dataset loaded. At
the file level, it currently supports zarr, single/multi-file HDF5 and
FITS files. The analysis functionality then available depends on the
dataset type. For example:

\begin{itemize}
\tightlist
\item
  Datasets which pair scattered pointsets with cataloged results of
  clustering algorithms have a natural association between such groups
  and their member points. Cosmological simulations in astrophysics
  provide one such motivating example, where halos and galaxies are
  identified in particle data. Scida then supports broadcasting of
  analysis and reduction operations over all groups for the associated
  particles.
\item
  Datasets containing spatial coordinates in one, two, or three
  dimensions can be queried for spatial subsets.
\item
  Datasets which are comprised of series, e.g.~snapshots at different
  times, or parameter variation suites, are automatically inferred and
  supported.
\end{itemize}

The dataset-dependent functionality of scida is handled with a flexible
and extensible mixin architecture.

\hypertarget{statement-of-need}{%
\section{Statement of need}\label{statement-of-need}}

Today, scientific ``big data'' is petabyte-scale. This makes traditional
analysis by a researcher on their personal computer prohibitive. Writing
analysis code which is distributed or out-of-core is complex, and is not
the main focus of scientists. The need for significant data management
creates (i) a barrier for new researchers, (ii) a substantial time
commitment apart from the science itself, and (iii) an increased risk of
errors and difficulties in reproducibility due to higher code
complexity, while (iv) workflows are often not easily transferable to
other datasets, nor (v) scalable and transferable to changing computing
resources.

\textbf{scida} solves these problems by providing a simple interface to
large datasets, hiding the complexity of the underlying data format and
file structures, and transparently handling the parallelization of
analysis computations. This is facilitated by the dask library, which
naturally separates the definition of a given computation from its
execution.

Initial support in scida is focused on astrophysical simulations and
observations, but the package is designed to be easily extensible to
other scientific domains. Existing analysis frameworks for astrophysical
simulations include python packages such as yt
(\protect\hyperlink{ref-yt}{Turk et al., 2011}), pynbody
(\protect\hyperlink{ref-pynbody}{Pontzen et al., 2013}), pygad
(\protect\hyperlink{ref-pygad}{Röttgers, 2018}), nbodykit
(\protect\hyperlink{ref-nbodykit}{Hand et al., 2018}) and swiftsimio
(\protect\hyperlink{ref-Borrow2021}{Borrow \& Kelly, 2021}). None
utilize the graph-based distributed analysis framework of dask. Often,
existing analysis packages rely on the explicit loading of entire
datasets into main memory. However, this approach is not transparently
scalable to large data sets, and requires the user to explicitly manage
the data chunks in custom analysis routines.

By providing a flexible interface to the dask library to handle large
data sets in a scalable fashion, users can also leverage dask
functionality and dask-based libraries such as dask-image
(\protect\hyperlink{ref-dask-image}{Kirkham \& others, 2018}) and
datashader (\protect\hyperlink{ref-datashader}{Bednar et al., 2022}).

scida was first utilized in Byrohl \& Nelson
(\protect\hyperlink{ref-Byrohl23}{2023}) for the analysis of
cosmological and radiative transfer simulations, particularly the
reduced chi-squared analysis exploring thousands of terabyte-sized
models.

\hypertarget{target-audience}{%
\section{Target Audience}\label{target-audience}}

\textbf{scida} aims to simplify access to large scientific data sets. It
lowers the barrier of entry for researchers to ask complex questions of
big data. As such, the scida package is targeted at researchers with
large data analysis problems. Its clean interface is appropriate for
users with no prior experience with big data or distributed data
analysis, as well as those who specifically want to leverage dask to
make their workflows easier to read and scalable.

Domain specific analysis routines can be implemented on top of scida.
Initial ``out of the box'' data support is currently focused on
astrophysical data sets, but scida aims to support other scientific
domains as well, where similar solutions will be beneficial.

\hypertarget{acknowledgements}{%
\section{Acknowledgements}\label{acknowledgements}}

CB and DN acknowledge funding from the Deutsche Forschungsgemeinschaft
(DFG) through an Emmy Noether Research Group (grant number NE 2441/1-1).

\hypertarget{references}{%
\section*{References}\label{references}}
\addcontentsline{toc}{section}{References}

\hypertarget{refs}{}
\begin{CSLReferences}{1}{0}
\leavevmode\vadjust pre{\hypertarget{ref-datashader}{}}%
Bednar, J. A., Crail, J., Crist-Harif, J., Rudiger, P., Brener, G., B,
C., Thomas, I., Mease, J., Signell, J., Liquet, M., Stevens, J.-L.,
Collins, B., Thorve, A., Bird, S., thuydotm, esc, kbowen, Abdennur, N.,
Smirnov, O., \ldots{} Bourbeau, J. (2022). \emph{Holoviz/datashader:
Version 0.14.3} (Version v0.14.3). Zenodo.
\url{https://doi.org/10.5281/zenodo.7331952}

\leavevmode\vadjust pre{\hypertarget{ref-Borrow2021}{}}%
Borrow, J., \& Kelly, A. J. (2021). \emph{Projecting SPH particles in
adaptive environments}. \url{https://arxiv.org/abs/2106.05281}

\leavevmode\vadjust pre{\hypertarget{ref-Byrohl23}{}}%
Byrohl, C., \& Nelson, D. (2023). The cosmic web in {Lyman-alpha}
emission. \emph{Monthly Notices of the Royal Astronomical Society},
\emph{523}, 5248--5273. \url{https://doi.org/10.1093/mnras/stad1779}

\leavevmode\vadjust pre{\hypertarget{ref-Dask}{}}%
Dask Development Team. (2016). \emph{Dask: {Library} for dynamic task
scheduling}.

\leavevmode\vadjust pre{\hypertarget{ref-Pint}{}}%
Grecco, H., \& others. (2012). Pint: Makes units easy. In \emph{GitHub
repository}. GitHub. \url{https://github.com/hgrecco/pint}

\leavevmode\vadjust pre{\hypertarget{ref-nbodykit}{}}%
Hand, N., Feng, Y., Beutler, F., Li, Y., Modi, C., Seljak, U., \&
Slepian, Z. (2018). {nbodykit: An Open-source, Massively Parallel
Toolkit for Large-scale Structure}. \emph{156}(4), 160.
\url{https://doi.org/10.3847/1538-3881/aadae0}

\leavevmode\vadjust pre{\hypertarget{ref-dask-image}{}}%
Kirkham, J., \& others. (2018). Dask-image: Distributed image
processing. \emph{GitHub Repository}.
\url{https://github.com/dask/dask-image}

\leavevmode\vadjust pre{\hypertarget{ref-pynbody}{}}%
Pontzen, A., Roškar, R., Stinson, G. S., Woods, R., Reed, D. M., Coles,
J., \& Quinn, T. R. (2013). \emph{{pynbody: Astrophysics Simulation
Analysis for Python}}.

\leavevmode\vadjust pre{\hypertarget{ref-pygad}{}}%
Röttgers, B. (2018). \emph{{pygad: Analyzing Gadget Simulations with
Python}} (p. ascl:1811.014). Astrophysics Source Code Library, record
ascl:1811.014.

\leavevmode\vadjust pre{\hypertarget{ref-yt}{}}%
Turk, M. J., Smith, B. D., Oishi, J. S., Skory, S., Skillman, S. W.,
Abel, T., \& Norman, M. L. (2011). {yt: A Multi-code Analysis Toolkit
for Astrophysical Simulation Data}. \emph{The Astrophysical Journal
Supplement Series}, \emph{192}, 9.
\url{https://doi.org/10.1088/0067-0049/192/1/9}

\end{CSLReferences}

\end{document}